\documentclass[aps,reprint,twocolumns,prl,groupeaddress,floatfix,notitlepage,nofootinbib]{revtex4-1}
\usepackage{amssymb,amsmath,amsfonts} 
\usepackage{mathtools}
\usepackage{physics}
\usepackage{array}
\usepackage{graphicx,epsfig,xcolor}
\usepackage[colorlinks=true,citecolor=blue,linkcolor=red]{hyperref}
\usepackage{newtxtext,newtxmath}
\begin{document}

\newcommand{\hatmath}[1]{\hat{\mathcal{#1}}} 

\title{Constant of motion identifying excited-state quantum phases}

\author{\'{A}ngel L. Corps}
   \email[]{angelo04@ucm.es}
    \affiliation{Departamento de Estructura de la Materia, F\'{i}sica T\'{e}rmica y Electr\'{o}nica \& Grupo Interdisciplinar de Sistemas Complejos (GISC), Universidad Complutense de Madrid, Av. Complutense s/n, E-28040 Madrid, Spain}
    
\author{Armando Rela\~{n}o}
\email[]{armando.relano@fis.ucm.es}
\affiliation{Departamento de Estructura de la Materia, F\'{i}sica T\'{e}rmica y Electr\'{o}nica \& Grupo Interdisciplinar de Sistemas Complejos (GISC), Universidad Complutense de Madrid, Av. Complutense s/n, E-28040 Madrid, Spain}

\date{\today} 

\begin{abstract}
We propose that a broad class of excited-state quantum phase transitions (ESQPTs) gives rise to two different excited-state quantum phases. These phases are identified by means of an operator, $\hatmath{C}$, which is a constant of motion only in one of them. Hence, the ESQPT critical energy splits the spectrum into one phase where the equilibirium expectation values of physical observables crucially depend on this constant of motion, and another phase where the energy is the only relevant thermodynamic magnitude. The trademark feature of this operator is that it has two different eigenvalues, $\pm1$, and therefore it acts as a discrete symmetry in the first of these two phases.  This scenario is observed in systems with and without an additional discrete symmetry; in the first case, $\hatmath{C}$ explains the change from degenerate doublets to non-degenerate eigenlevels upon crossing the critical line. We present stringent numerical evidence in the Rabi and Dicke models, 
suggesting that this result is exact in the thermodynamic limit, with finite-size corrections that decrease as a power-law.  
\end{abstract}

\maketitle

\textit{Introduction.-} A quantum phase transition happens when an abrupt change in the
ground state of a physical system is observed. The corresponding
critical point is signaled by a non-analyticity, and two different
phases can be identified by equilibrium measurements
\cite{Sachdev}. Excited-state quantum phase transitions (ESQPTs), a generalization of this phenomenon to excited states,
have been the focus of intense research during the last
years \cite{Caprio2008,Stransky2014} (for a recent, excellent review, see \cite{Cejnar2020}). ESQPTs give rise to a great variety of dynamical
consequences, like huge decoherence \cite{Relano2008,Perez2009};
singularities in quench dynamics \cite{Perez2011,Santos2015,Lobez2016,Bernal2017,Kloc2018}, feedback
control in dissipative systems \cite{Kopylov2015}, quantum work
statistics \cite{Wang2017}, and localization \cite{Santos2016};
symmetry-breaking equilibrium states
\cite{Puebla2013,Puebla2014}; universal dynamical scaling
\cite{Puebla2020}; dynamical instabilities \cite{Bastidas2014};
irreversibility without energy dissipation \cite{Puebla2015}, and
reversible quantum information spreading \cite{Hummel2019}. They are
somehow linked to thermal phase transitions
\cite{Bastarrachea2016,Perez2017} and dynamical phase transitions
\cite{Puebla2020PRB,Cabedoarxiv}. They can be identified by their
consequences in the classical \cite{Feldmann2020} and semi-classical
\cite{Wang2020,Klocarxiv} phase-space dynamics \cite{Wang2020arxiv,Wang2019}, as in the singularities of the density of states \cite{Stransky2016}. Its signatures have been theoretically and
experimentally observed in several physical systems
\cite{Zobov2005,Larese2013,Dietz2013,Tian2020,Khalouf2021,Larese2011}, and its connections with quantum Lyapunov exponents have been explored \cite{Quiroz-Juarez2020,Chavez-Carlos2019,Bastarrachea-Magnani2016}. However, no physical features of standard phase transitions have been identified yet.

In this Letter, we show that a typical feature of a large class of ESQPTs splits the spectrum into two different excited-state quantum phases. These are identified by an operator which is a constant of motion in just one of them (usually, below the ESQPT). This constant of motion signals to which part of the semiclassical phase space a given quantum state is attached to, and assigns it a conserved quantum number, with important thermodynamic consequences. We present general arguments, and illustrate our findings with the paradigmatic Rabi (RM) \cite{Rabi1936,Rabi1937} and the Dicke (DM) \cite{Dicke1954,Bastarrachea2014,Bastarrachea2014II,Relano2016,Bastarrachea2017} models, discussing dynamical and thermodynamic consequences.

\textit{Constant of motion.-} We start from the classical limit of a quantum system with Hamiltonian ${ H}(\mathbf{x})$, where $\mathbf{x}\in\mathbb{R}^{2\nu}$ accounts for all relevant canonical coordinates, and $\nu\in\mathbb{N}$ is the number of classical degrees of freedom. ESQPTs are caused by fixed points, $\mathbf{x}_{c}$, of the classical Hamiltonian flow, $\nabla {H} ({\mathbf x_c})=0$, at a critical energy $E_c \equiv { H} (\mathbf{x_c})$ \cite{Cejnar2020}. The main result of this Letter is the following conjecture:

{\em Suppose there exists a dynamical function, $f({\mathbf x})$, satisfying the following two properties: \emph{(i)} $f(\mathbf{x_c})=0$; and \emph{(ii)} on one side of the transition (say $E<E_c$), every trajectory verifies either $f(t) \equiv f({\mathbf x}[t]) <0$ or $f(t)>0$, $\forall t$, depending on the initial condition, but this is no longer true on the other side. Then, there exists a quantum operator,} $\hatmath{C}\equiv \textrm{sign}\,[\hat{f}(\mathbf{ \hat{x}})]${\em, which is a constant of motion only in the first of these two phases, $E<E_c$.}

Here, the sign of an operator $\hat{f}$ is defined  $\textrm{sign}\,(\hat{f})\equiv F\,\textrm{sign}\,(D)\,F^{-1}=F\,\textrm{diag}\,[\textrm{sign}\,(\{d_{i}\})]F^{-1}$, where $D$ is a diagonal matrix whose elements $\{d_{i}\}_{i}$ are the eigenvalues of $\hat{f}$, and $F$ is a matrix whose columns are the eigenvectors of $\hat{f}$. Hence, the operator ${\hatmath C}$ has only two eigenvalues, $\pm 1$, and therefore represents a $\mathbb{Z}_2$ symmetry in this phase. Notwithstanding, it is unrelated to any exact discrete symmetry of a given model, and thus it is not linked to spontaneous symmetry-breaking observed in some phase transitions. As representative examples of systems fulfilling the conditions for the above conjecture, we quote the Lipkin-Meshkov-Glick model, the Rabi, and Dicke models, spinor Bose-Einstein condensates and Bose mixtures in a double-well potential, the coupled top, and the two-fluid Lipkin model \cite{Caprio2008,Relano2008,Perez2009,Kopylov2015,Puebla2013,Perez2017,Feldmann2020,Wang2020,Klocarxiv,Wang2020arxiv,Bastarrachea2014,Brandes2013,Puebla2016,Relano2014,Garcia-Ramos2017}. In all these systems, there is an ESQPT at $E=E_{c}$ below which the classical phase space is split into disconnected wells. Then, $\hatmath{C}$ indicates to which classical well a quantum state belongs. 

To get a more precise definition of these excited-state quantum phases, we write the quantum Hamiltonian $\hat{\mathcal{H}}=\sum_n E_n \hat{P}_n$, where $\hat{P}_n$ is the projector onto the eigenspace with energy $E_n$. Thus, $\left[\hat{\mathcal{C}},\hat{P}_n\right]=0$, $\forall n \, / \, E_n < E_c$, and $\left[\hat{\mathcal{C}},\hat{P}_n\right] \neq 0$, $\forall n \, / \, E_n > E_c$. This means that $\langle \hatmath{C} \rangle$ is conserved by any-time evolution verifying that $\langle \hatmath{H} \rangle =E<E_c$, but this conservation rule no longer holds if $E>E_c$. Hence, there exists a phase in which the expectation value of this observable must be taken into account to properly describe both equilibrium \cite{Guryanova2016} and non-equilibrium \cite{Mur2018} thermodynamics. By contrast, the other phase is characterized by standard thermodynamics.

\textit{Numerical test.-} As an illustration, we perform a numerical test on a generalization of the RM and the DM. Both models account for the interaction between a monochromatic bosonic field and $N$ identical two-level atoms. The Hamiltonian reads
\begin{equation}\label{eq:model}
\hat{\mathcal{H}}_{\alpha}=\omega\hat{a}^{\dagger}\hat{a}+\omega_{0} \hat{J}_{z}+\frac{2\lambda}{\sqrt{N}}(\hat{a}^{\dagger}+\hat{a})\hat{J}_{x}+\sqrt{\frac{N\omega_{0}}{2}}\alpha(\hat{a}^{\dagger}+\hat{a}),
\end{equation}
where $\hat{a}^{\dagger}$ and $\hat{a}$ are the appropriate bosonic creation and annihilation operators, 
and $\mathbf{\hat{J}}=(\hat{J}_{x},\hat{J}_{y},\hat{J}_{z})$ is an angular momentum. The total angular momentum $\mathbf{\hat{J}}^{2}$ is conserved.  The dynamics of a set of $N$ identical two-level atoms is recovered with $j=N/2$, which we use.
Then, $\omega_0$ represents the constant splitting of the atom eigenlevels, while $\omega$ represents the frequency of the photons to which atoms are coupled by the parameter $\lambda$.

The case $\alpha=0$ is the standard RM and DM. The Hamiltonian $\hat{\mathcal{H}}_{\alpha=0}$ has a discrete $\mathbb{Z}_{2}$  symmetry, called \textit{parity}, allowing to separate eigenstates according to $\hat{\Pi}\ket{E_{n,\pm}}=\pm\ket{E_{n,\pm}}$ where $\hat{\Pi}\equiv \exp[i\pi(j+\hat{J}_{z}+\hat{a}^{\dagger}\hat{a})]$ and $\hat{\mathcal{H}}_{0}\ket{E_{n,\pm}}=E_{n,\pm}\ket{E_{n,\pm}}$.
 If $\alpha\neq0$, $\hat{\Pi}$ is not conserved, $[\hat{\mathcal{H}}_{\alpha\neq0},\hat{\Pi}]\neq0$. 

Equation \eqref{eq:model} admits two different thermodynamic limits (TLs): (i) in the DM, the number of two-level atoms goes to infinity, $N\to\infty$, fixing $\omega_{0}/\omega<\infty$; (ii) in the RM, $\omega_0/\omega \rightarrow \infty$, fixing $N=1$ \cite{Puebla2016}. Thus, in the RM (DM) we set $N=1$ ($\omega=\omega_{0}=1$) and let $\omega_{0}/\omega$ ($N$) be the scaling parameter, which we denote simply $\mathcal{N}$. We use the reduced energy scale $\epsilon\equiv E/(\omega_{0}j)$. In both models, there exists a certain coupling $\lambda_{c}$ only above which ESQPTs start appearing at a certain critical energy \cite{sm}. 

\begin{center}
\begin{figure}
\hspace*{-0.60cm}\includegraphics[width=0.54\textwidth]{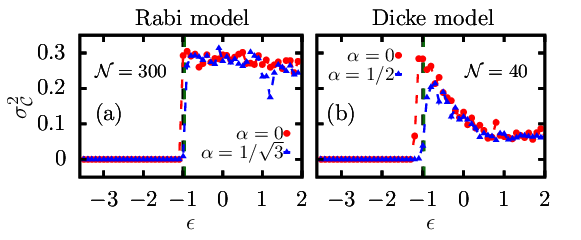}
\caption{Variance of the time-evolution $\langle \hatmath{C}(t)\rangle$ as a function of energy with $\lambda=3\sqrt{\omega\omega_{0}}/2>\lambda_{c}$ for the RM (a) and DM (b). Black (green) dashed lines mark the ESQPT critical energy for $\alpha=0$ ($\alpha=1/\sqrt{3}$ in the RM and $\alpha=1/2$ in the DM) \cite{sm}. }
\label{panelsigma}
\end{figure}
\end{center}

\begin{center}
\begin{figure}
\hspace*{-1cm}\includegraphics[width=0.59\textwidth]{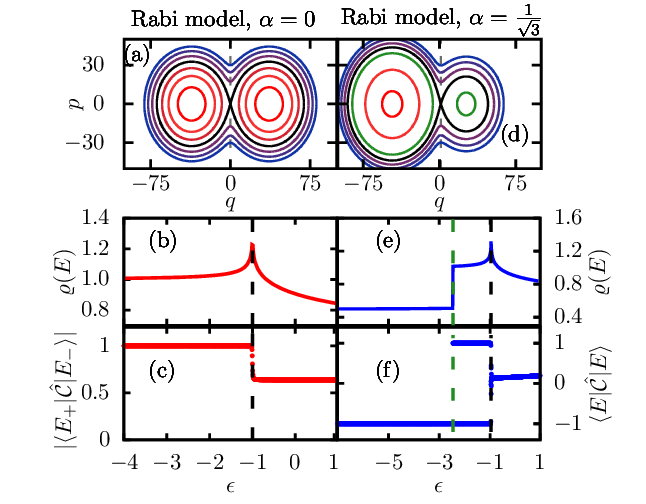}
\caption{(a-c) Semiclassical phase space, density of states and expectation values of $\hat{\mathcal{C}}$ in quantum eigenstates for $\alpha=0$ ($\mathcal{N}=300,\,\lambda=3\sqrt{\omega\omega_{0}}/2>\lambda_{c}$) in the RM. (d-f) Same for $\alpha=1/\sqrt{3}$. The vertical dashed lines mark the critical energies $\epsilon_{c}(\alpha=0)=-1$ and $\epsilon_{c}(\alpha=1/\sqrt{3})\approx -0.96$ (black) and $\epsilon_{c_{2}}(\alpha=1/\sqrt{3})\approx -2.46$ (green).  If $\alpha=0$, both wells are symmetric and at the same energy. If $\alpha\neq0$, the second energy well is inaccessible if $\epsilon<\epsilon_{c_{2}}$. Black curves in (a,d) correspond to the ESQPT energy connecting both wells.}
\label{fig:comparaciones}
\end{figure}
\end{center}

These models fulfill the condition for the existence of the operator $\hat{\mathcal{C}}$. From their semiclassical energy surfaces \cite{Puebla2013,Bastarrachea2014,Bastarrachea2014II,Relano2016,Bastarrachea2017,sm} [Fig. \ref{fig:comparaciones}(a,d)] we propose that the relevant dynamical function is $f(\mathbf{x})=q-q_{c}(\alpha,\lambda)$, where $q_{c}(\alpha,\lambda)$ is the canonical coordinate corresponding to the ESQPT critical energy.
If $\alpha=0$, $q_{c}(0,\lambda)=0$, $\forall \lambda>\lambda_{c}$; if $\alpha\neq0$, $q_{c}(\alpha,\lambda)$ is a complicated function \cite{sm}. Hence, $\hatmath{C}$ takes the form

\begin{equation}\label{eq:constant} \hat{\mathcal{C}}= \textrm{sign}\,\left[\hat{q}-q_{c}(\alpha,\lambda)\right],
\end{equation}
 where $\hat{q}=(\hat{a}^{\dagger}+\hat{a})/\sqrt{2}$ in the RM and $\hat{q}=(\hat{a}^{\dagger}+\hat{a})/\sqrt{2j}$ in the DM.
 
 To test this hypothesis, we work with $\lambda={3}\sqrt{\omega\omega_{0}}/2$, and two different values of the perturbation, $\alpha$; in both cases, $\lambda>\lambda_{c}$, so that ESQPTs exist. We chose an initial state, $\ket{\Psi(t=0)}$, with 10 consecutive eigenstates equally populated. Then, we calculate the time evolution $\langle\hat{\mathcal{C}}(t)\rangle=\bra{\Psi(t)}\hat{\mathcal{C}}\ket{\Psi(t)}$ for $M=100$ time steps. Next, we calculate its mean, accounting for the long-time average $\overline{\langle\hat{\mathcal{C}}\rangle}=(1/M)\sum_{i=1}^{M}\langle\hat{\mathcal{C}}(t_{i})\rangle$. Finally, we account for the fluctuations around this value using the width
$\sigma_{\mathcal{C}}^{2}=(1/M)\sum_{i=1}^{M}\left(\langle\hat{\mathcal{C}}(t_{i})\rangle-\overline{\langle{\hat{\mathcal{C}}}\rangle}\right)^{2}$.
This protocol is repeated for different initial states with increasing energy values, probing different regions of the spectrum. 
Results are shown in Fig. \ref{panelsigma}. In the RM, we work with $\mathcal{N}=300$ and $\alpha\in\{0,1/\sqrt{3}\}$, while in the DM $\mathcal{N}=40$ and $\alpha\in\{0,1/2\}$. We can see that $\sigma^2_{\mathcal{C}}$ jumps abruptly from $\sigma^2_{\mathcal{C}}=0$ to $\sigma^2_{\mathcal{C}}>0$ at the corresponding ESQPT criticalities in the RM for both values of $\alpha$. This means that $\hatmath{C}$ stops being constant at this critical energy.
The jump is not so abrupt in the DM because $\mathcal{N}$ is one order of magnitude smaller. The different behavior of $\sigma_{\mathcal{C}}^{2}$ abvove the ESQPT in the RM and DM is presumably linked to their different classical degrees of freedom; however, this does not affect the results of this Letter \cite{sm}.

{\em Properties of the low-energy phase.-} Our next step is to derive some mathematical consequences of the previous facts. We begin with the case $\alpha \neq 0$. Fig. \ref{fig:comparaciones}(e) shows the semiclassical density of states, $\varrho(\epsilon)\equiv \frac{1}{\mathcal{N}}\frac{1}{(2\pi)^{\nu}}\int\textrm{d}^{\nu}p\int \textrm{d}^{\nu}q\,\delta[\epsilon-H_{\alpha}(p,q)]$, for the RM with $\alpha=1/\sqrt{3}$. We observe two ESQPTs, marked by (i) a logarithmic singularity at $\epsilon_c \approx -0.96$, and (ii) a finite jump at $\epsilon_{c_2}\approx-2.46$. Fig. \ref{fig:comparaciones}(d) shows the corresponding semiclassical phase space. Below $\epsilon_{c_{2}}$, there is a single connected region of constant energy curves, on the left. At $\epsilon_{c_{2}}$, a disconnected second region appears on the right. At $\epsilon_{c}$, both become connected. 

Fig. \ref{fig:comparaciones}(f) shows the quantum diagonal expectation values of $\hat{\mathcal{C}}$. Our conjecture applies for $\epsilon < \epsilon_c$: in this region, eigenstates of the Hamiltonian are also eigenstates of $\hatmath{C}$, and therefore $\bra{E_{n}}\hat{\mathcal{C}}\ket{E_{n}}=\pm 1$. But this operator also provides a description for the other ESQPT. Below $\epsilon_{c_2}$ all the eigenstates verify $\bra{E_{n}}\hat{\mathcal{C}}\ket{E_{n}}=- 1$. This means that $\hatmath{C}$ is not necessary to account for the physics of observables in equilibrium: any initial state is characterized by $\left< \hatmath{C} \right>=-1$ within this region, as all quantum eigenstates are attached to the left well of the classical phase space. By contrast, Fig. \ref{fig:comparaciones}(f) also clearly shows that the conditions for the eigenstate thermalization hypothesis (ETH) \cite{Alessio2016,Tasaki1998,Rigol2008,Reimann2015,Deutsch2018,Srednicki1994} are not fulfilled if $\epsilon_{c_2} < \epsilon < \epsilon_c$ --- $\bra{E_{n}}\hat{\mathcal{C}}\ket{E_{n}}$ jumps abruptly from $-1$ to $1$. This is because quantum eigenstates can belong either to the left or to the right well in this region. Therefore, \textit{the energy is not enough to describe equilibrium thermodynamics}, and additional information, given by the knowledge of $\left< \hatmath{C} \right>$, is also required.

If $\alpha=0$, there is an additional $\mathbb{Z}_{2}$ symmetry, $[\hat{\mathcal{H}}_{\alpha=0},\hat{\Pi}]=0$. It can be easily shown that $\hat{\mathcal{C}}$ changes the parity  of any Fock state with well-defined parity \cite{sm}. 
This means that $\hat{\Pi}$ and $\hatmath{C}$ cannot be diagonalized in the same basis as they do not commute, $[\hatmath{C},\hat{\Pi}]\neq0$. Hence, as $\left[\hat{P}_n, \hat{\Pi} \right]=\left[\hat{P}_n, \hatmath{C} \right]=0$ for every energy subspace with $E_n<E_c$, there exists two different bases diagonalizing this part of the Hamiltonian, and thus {\em all energy levels in this excited-state phase must be doubly degenerate}. By contrast, $\hat{\Pi}$ is the only discrete $\mathbb{Z}_2$ symmetry if $E>E_c$, and therefore energy levels are not expected to be degenerate in that phase. This phenomenology has been observed in a large number of models displaying ESQPTs \cite{Cejnar2020,Relano2008,Perez2009,Sindelka2017,Cejnar2006,Cejnar2007,Ribeiro2008,Relano2014,Garcia-Ramos2017,Rivera2021}. As a consequence, the eigenvectors of $\hatmath{C}$ are $ \left(\ket{E_{n,+}}\pm \ket{E_{n,-}}\right)/\sqrt{2}$, and its expectation values are $\left< E_{n,-} \right| \hatmath{C} \left| E_{n,+} \right>=\pm 1$, due to an arbitrary sign coming from the relative phase between the degenerate eigenstates $\ket{E_{n,+}}$ and $\ket{E_{n,-}}$ \cite{sm}. This prediction is illustrated in Fig. \ref{fig:comparaciones}(a-c). The density of states in Fig. \ref{fig:comparaciones}(b) shows that there is a single ESQPT at $\epsilon_c=-1$ in this case.  It marks the energy at which the two equivalent wells in the classical phase space, Fig. \ref{fig:comparaciones}(a), become connected; below $\epsilon_{c}$, we have two disconnected, symmetric wells. Fig. \ref{fig:comparaciones}(c) clearly shows that our conjecture indeed holds below this energy: $\left|\bra{E_{n,-}}\hatmath{C}\ket{E_{n,+}}\right|=1$ for $E<E_c$, whereas $\left|\bra{E_{n,-}}\hatmath{C}\ket{E_{n,+}}\right |\neq 1$ for $E>E_c$.

\begin{center}
\begin{figure}[h!]
\hspace*{-1.4cm}%
\includegraphics[width=0.6\textwidth]{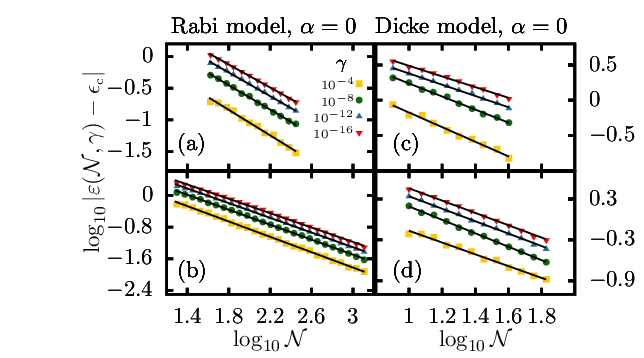}
\caption{Finite-size scaling of the ESQPT precursor $\varepsilon(\mathcal{N},\gamma)$ for different bounds $\gamma$ (points), extracted from (a,c) the expectation values of $\hat{\mathcal{C}}$ and (b,d) eigenlevel degeneracies \cite{sm}. Solid lines represent the best linear fit to the points.}
\label{scalings}
\end{figure}
\end{center}

\textit{Finite-size scaling.-} Physically, this last case is more challenging since we need two non-commuting discrete symmetries, $\hat{\Pi}$ and $\hatmath{C}$, to build a complete description of thermodynamic equilibrium. Hence, we work with $\alpha=0$, which has been the object of recent experimental works \cite{Cai2021,Baumann2011}, to perform a stringent test on our conjecture. As explained above, it implies energy doublets, $|E_{n,+}-E_{n,-}|=0$, and also $\left|\bra{E_{n,+}}\hat{\mathcal{C}}\ket{E_{n,-}}\right|=1$ if $E_{n}<E_{c}$, in the TL. To study finite systems, we define a finite-size precursor of the ESQPT as the energy, $\varepsilon(\mathcal{N},\gamma)$, above which the gap $\Delta E_{n}\equiv |E_{n,+}-E_{n,-}|/\langle s\rangle >\gamma$ or the difference $1-\left|\bra{E_{n,+}}\hat{\mathcal{C}}\ket{E_{n,-}}\right|>\gamma$ for a given small bound $\gamma>0$. Here, $\langle s\rangle$ is the mean level spacing calculated within a window of 10 eigenlevels around the target energy. According to our conjecture, $\lim_{\mathcal{N}\to\infty}\varepsilon(\mathcal{N},\gamma)=\epsilon_{c}$ for any (small) $\gamma$; for finite systems, the closer the energy to the ESQPT, the larger $\mathcal{N}$ is needed to get the above indicators below a given bound \cite{sm}. In Fig. \ref{scalings}(a,c) we test this result with $\varepsilon(\mathcal{N},\gamma)$ extracted from the condition on $\left|\bra{E_{n,+}}\hat{\mathcal{C}}\ket{E_{n,-}}\right|$; in Fig. \ref{scalings}(b,d), $\varepsilon(\mathcal{N},\gamma)$ is obtained from the condition on $\Delta E_{n}$ instead.  We can clearly see that $\left|\varepsilon(\mathcal{N},\gamma)-\epsilon_{c}\right|\propto \mathcal{N}^{-\beta}$, with $\beta>0$, in all cases, with bounds ranging from $\gamma=10^{-16}$ to $\gamma=10^{-4}$. This strongly suggests that, in the TL, $\varepsilon(\mathcal{N},\gamma)\to\epsilon_{c}$ following a power-law in $\mathcal{N}$. In other words, the finite-size scaling around $\epsilon_{c}$ behaves like in standard quantum and thermal phase transitions.

\textit{Thermodynamic consequences.-} Finally, we tackle the thermodynamic implications by studying the properties of observables in equilibrium in the DM with $\alpha=0$ (similar results are obtained in the RM).
We start from its most general ground-state at $\lambda_{\textrm{ini}}=\frac{3}{2}\lambda_{c}>\lambda_{c}$, given by $\ket{\Psi(t=0)}=\sqrt{p}\ket{E_{\textrm{GS},+}}+e^{i\phi}\sqrt{1-p}\ket{E_{\textrm{GS},-}}$,
where $\hatmath{H}_{0}(\lambda_{\textrm{ini}})\ket{E_{\textrm{GS},\pm}}=E_{\textrm{GS}}\ket{E_{\textrm{GS},\pm}}$, $0\leq p\leq 1$ 
and $0\leq\phi<2\pi$. In this state, $\bra{\Psi(0)}\hatmath{C}\ket{\Psi(0)}=2\sqrt{p(1-p)}\cos\phi.$ We then perform a quench, $\lambda_{\textrm{ini}}\to\lambda_{\textrm{fin}}$, so the time-evolved density matrix $\hat{\rho}(t) \equiv \ket{\Psi(t)}\bra{\Psi(t)}$ reads ($\hbar=1$)
\begin{equation}
\hat{\rho}(t)= \sum_{m,n} \sum_{k,\ell=\pm} c_{m,k}c^*_{n,\ell} \text{e}^{-i(E_{m,k} - E_{n,\ell})t} \ket{E_{m,k}} \bra{E_{n,\ell}},
\end{equation}
where $c_{m,k} \equiv \braket{E_{m,k}}{\Psi(0)}$ \cite{sm}, and all the eigenenergies and eigenstates after the quench are those of ${\mathcal H}_{0}(\lambda_\textrm{fin})$, $\hat{\mathcal{H}}_{0}(\lambda_{\textrm{fin}})\ket{E_{m,k}}=E_{m,k}\ket{E_{m,k}}$. 

First, we consider the time-evolved expectation value $\left< {\hatmath C(t)}\right>\equiv \textrm{Tr} \left[ \hat{\rho}(t) {\hatmath C}\right]$. From Fig. \ref{scalings}(a,c) we can conclude that only the terms with
$m=n$ and $k=-\ell$ contribute to $\left< {\hatmath C(t)}\right>$ in the TL, since $\langle E_{m,k}|\hat{\mathcal{C}}|E_{n,\ell}\rangle=\pm\delta_{m,n}(1-\delta_{k,\ell})$ if $E_{m},E_{n} < E_c$; and from Fig. \ref{scalings}(b,d) we also conclude that this contribution always remains constant and
depends exclusively on the initial condition encoded in  $c_{m,k}$,     $\langle\hat{\mathcal{C}}(t)\rangle=\sum_{n}c_{n,+}^{*}c_{n,-}\langle E_{n,+}|\hat{\mathcal{C}}|E_{n,-}\rangle$. Hence, \textit{the abrupt change inferred from Fig. \ref{scalings} implies a change from constant to non-constant $\langle\hat{\mathcal{C}}(t)\rangle$ at the critical energy in the TL} \cite{sm}.

A further consequence is that 
the long-time average of the time-evolved wavefunction in this phase \cite{Reimann2008}, $\overline{\hat{\rho}}=\lim_{\tau \rightarrow \infty} (1/\tau) \int_0^{\tau} \textrm{d} t \hat{\rho}(t)$, generally depends on both $\langle\hat{\Pi}\rangle$ and $\langle\hat{\mathcal{C}}\rangle$ ---not only on the energy as would be expected in the microcanonical ensemble. The long-time average of a typical observable $\hat{\mathcal{O}}$ is therefore $\overline{\langle \hatmath{O}\rangle}\equiv\lim_{\tau\to\infty}(1/\tau)\int_{0}^{\tau}\textrm{d}t\,\bra{\Psi(t)}\hatmath{O}\ket{\Psi(t)}= \textrm{Tr}[\overline{\hat{\rho}}\hatmath{O}]$. 
To explore this, we choose $\lambda_{\textrm{fin}}=3\lambda_{c}$; then the energy of the non-equilibrium state is $\epsilon_{\textrm{fin}}=\bra{\Psi(0)}\hatmath{H}_{0}(\lambda_{\textrm{fin}})\ket{\Psi(0)}/(\omega_{0}j)\approx-3.15<\epsilon_{c}=-1$. We monitor the behavior of two representative observables:  $\hat{J}_{x}$ and $\mathcal{S}(t)$, after letting the system relax during a time $t=10^6$ $\mu$s, by means of $\tau=10^3$ equal steps, considering the realization of the DM discussed in \cite{Mur2018}. Here, $\mathcal{S}(t)$ is the entanglement entropy, $\mathcal{S}(t)\equiv -\Tr[\hat{\rho}_{s}(t)\log\hat{\rho}_{s}(t)]$, where $\hat{\rho}_{s}(t)=\Tr_{E}\ket{\Psi(t)}\bra{\Psi(t)}$, with the `environment' corresponding to the photonic radiation and the `system' being the atomic part of the Hamiltonian.

\begin{center}
\begin{figure}[h!]
\hspace*{-0.55cm}\includegraphics[width=0.56\textwidth]{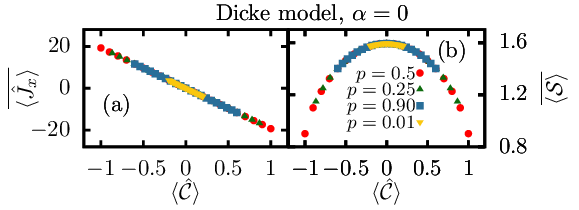}
\caption{Long-time averages of (a) $\hat{J}_{x}$ and (b) $\mathcal{S}$ following a quench $\lambda_{\textrm{ini}}=\frac{3}{2}\lambda_{c}\to\lambda_{\textrm{fin}}=3\lambda_{c}$ in the DM ($\alpha=0$) from initial states with different $p$ and $\phi$.}
\label{cphi}
\end{figure}
\end{center}

We observe in Fig. \ref{cphi} that both long-time averages $\overline{\langle \hat{J}_x\rangle}$ and $\overline{\langle \mathcal{S}\rangle} $ crucially depend on $\langle  \hatmath{C} \rangle=\overline{\langle\hatmath{C}\rangle}$. This means that the system reaches different equilibrium states characterized by the same energy, depending on the initial value of the coherence between parity sectors, given by the angle $\phi$. 
Therefore, \textit{one must take into account the expected value of $\hatmath{C}$ to properly describe equilibrium states below the critical energy of the ESQPT,} much in the same way that in the region $\epsilon_{c_2} < \epsilon < \epsilon_c$ in the case with $\alpha \neq 0$.  We note that neither $\overline{\langle \hat{J}_{x}\rangle}$ nor $\overline{\langle\mathcal{S}\rangle}$ depend on $p$. Thus, they do not depend on $\langle \hat{\Pi} \rangle$  either, even though parity is an exact constant of motion in both phases. Therefore, we conclude that the role played by $\hatmath{C}$ in thermodynamics is much more important than that of $\hat{\Pi}$.

Finally, we remark that after a quench onto the normal phase, long-time averages only depend on the final energy $\epsilon_{\textrm{fin}}$ (not shown), as expected in the standard microcanonical ensemble \cite{Alessio2016}.

\textit{Conclusions.-} The main result of this Letter can be summarized in the following phase diagram characterizing two phases with different dynamical and thermodynamic properties:

\textbullet \, A phase where there exists an operator, $\hat{\mathcal{C}}$, with two eigenvalues, $\textrm{Spec}( \hat{\mathcal{C}})=\{\pm 1\}$, commuting with the corresponding part of the Hamiltonian, $[\hatmath{C},\hat{P}_{n}]=0$, $E_{n}<E_{c}$.
If the system has a discrete symmetry and $[\hat{\mathcal{C}},\hat{\Pi}]\neq0$, then this is a broken-symmetry phase where equilibrium states are a mixture of states with broken $\hat{\Pi}$, broken $\hat{\mathcal{C}}$, or both. Furthermore, thermodynamics crucially depends on $\langle \hatmath{C} \rangle $. 

\textbullet \, A normal phase, where $\hat{\mathcal{C}}$ is no longer a constant of motion. If there exists a discrete $\mathbb{Z}_2$ symmetry, then all the eigenstates of the Hamiltonian are also eigenstates of this symmetry in this phase.

Additionally, $\hatmath{C}$ provides a description of other kind of ESQPTs too, as exemplified by the abrupt jump in the level density of the RM and DM with $\alpha \neq 0$, and versions of the Lipkin model \cite{Perez2009,Ribeiro2008,Lipkin1965}.

This Letter provides a powerful framework to identify a broad class of ESQPTs dynamically, as the number of constants of motion abruptly changes at the corresponding critical energy. This should entail important consequences for non-equilibrium processes crossing an ESQPT due to the change of conserved charges \cite{Mur2018,Puebla2013}, and also for the steady states resulting from dynamical phase transitions \cite{Puebla2020PRB,Lewis2021}. A consequence of the non-commutativity of $\hatmath{C}$ and $\hat{\Pi}$ is the possibility to build equilibrium states in which the information about both the population of each symmetric well and the quantum coherence between them is recorded.

Further research is needed to determine how to link our results with typical features of standard phase transitions, like critical slowing down. Experimental tests involving broken-symmetry equilibirium states \cite{Cai2021,Baumann2011} will play a relevant role in this endeavor.

\begin{acknowledgments}
We acknowledge financial support from Ministerio de Ciencia, Innovaci\'{o}n y Universidades/Agencia Estatal de Investigaci\'{o}n (MCIU/AEI/FEDER, EU) under Grant 
No. PGC2018-094180-B-I00.
\end{acknowledgments}

\end{document}


\newcommand{\hatmath}[1]{\hat{\mathcal{#1}}} 

\title{Supplemental Material for: \\Constant of motion identifying excited-state quantum phases}

\author{\'{A}ngel L. Corps}
   \email[A. L. Corps: ]{angelo04@ucm.es}
    \affiliation{Departamento de Estructura de la Materia, F\'{i}sica T\'{e}rmica y Electr\'{o}nica \& Grupo Interdisciplinar de Sistemas Complejos (GISC), Universidad Complutense de Madrid, Av. Complutense s/n, E-28040 Madrid, Spain}
    
\author{Armando Rela\~{n}o}
\email[A. Rela\~{n}o: ]{armando.relano@fis.ucm.es}
\affiliation{Departamento de Estructura de la Materia, F\'{i}sica T\'{e}rmica y Electr\'{o}nica \& Grupo Interdisciplinar de Sistemas Complejos (GISC), Universidad Complutense de Madrid, Av. Complutense s/n, E-28040 Madrid, Spain}

\date{\today} 

\maketitle

In this Supplemental Material, (i) we review the classical energy curves of the Rabi (RM) and Dicke (DM) models, (ii) we compare $\sigma_{\mathcal{C}}^{2}$ of Fig. 1 with a smaller system size, (iii) we show that $\hat{\mathcal{C}}$ changes parity of any Fock state, detailing the mathematical calculation of the eigenstates and eigenvalues of $\hat{\mathcal{C}}$ in the RM and DM ($\alpha=0$), (iv) we give additional figures regarding the finite-size scaling of Fig. 3, (v) we provide examples of the time evolution of $\hat{\mathcal{C}}$ below and above the ESQPT, and (vi) we give explicit information about the initial states used and represent the final eigenstate population after a typical quench.

\section{S1. Semiclassical constant energy surfaces}
 
 As indicated in the main text, the models that we use to illustrate our findings, the Rabi ($N=1$) and Dicke models $(N\geq2)$, have two different thermodynamic limits (TL). First is the standard TL: the number of two-level atoms goes to infinity, $N\to\infty$, keeping fixed $\omega_{0}/\omega<\infty$. This limit gives rise to a classical Hamiltonian with $\nu=2$ degrees of freedom, showing both a ground-state quantum phase transition (QPT) \cite{Emary2003,Emary2003b} and an ESQPT \cite{Perez2011,Perez2011b,Brandes2013}. Second is the limit $\omega_0/\omega \rightarrow \infty$, keeping fixed $N<\infty$.  Remarkably, this limit makes it possible to observe both a QPT and an ESQPT in the RM \cite{Hwang2015,Puebla2016}, even though it has just one atom. It is well described by a classical model with only $\nu=1$ degree of freedom. If $\alpha=0$, both the RM and DM display a second-order QPT at $\lambda_{c}\equiv \sqrt{\omega\omega_{0}}/2$. For $\lambda<\lambda_{c}$, the system is in the normal phase, while if $\lambda>\lambda_{c}$ it is in the superradiant phase, where it also exhibits an ESQPT at the critical line $\epsilon_{c}(\alpha=0)=-1$ in the TL. If $\alpha\neq 0$, both $\epsilon_{{c}}$ and the critical coupling above which there appears ESQPTs become involved functions of $\lambda$ itself and $\alpha$. 
 
 The classical energy surfaces of these two models are reviewed below. For the RM, the semiclassical model can be understood as a precession of a very large collective spin, accounting for the bosonic field. In the DM, we have two coupled large spins instead, one for the set of $N$ atoms, and another for the bosonic field. 
 
\subsection{Rabi model}
To begin with, we focus on the RM, with a single effective degree of freedom ($\nu=1$). To obtain a more general model allowing for asymmetrical and non-equivalent minima, we add a perturbation term $\hat{V}_{\alpha}=\sqrt{N\omega_{0}/2}\alpha(\hat{a}^{\dagger}+\hat{a})$ to the standard Rabi Hamiltonian, as in the main text, Eq. (2). In the RM, $N=1$. Then it follows directly from Refs.  \cite{Puebla2016,Relano2016,Bastarrachea2017} that the constant energy curves of the semiclassical analogue follow the expression
\begin{equation}
\label{eq:clasicoRM}
    H_{\alpha}(p,q)\equiv\frac{\widetilde{H}_{\alpha}(p,q)}{\omega_{0}j}=\frac{\omega}{\omega_{0}} \left(p^2+q^2 \right) - \sqrt{1+\frac{\lambda^2 \omega q^{2}}{\lambda_c^2 \omega_0 j}}+\frac{\alpha q}{j\sqrt{\omega_{0}}},
\end{equation}
where $j=1/2$ and $\lambda_{c}=\sqrt{\omega\omega_{0}}/2$.  The semiclassical phase space is therefore two-dimensional, $\mathcal{M}\subseteq\mathbb{R}^{2}$, and its canonical variables $\mathbf{x}=(p,q)\in\mathbb{R}^{2}$ are $p=(i/\sqrt{2})(\hat{a}^{\dagger}-\hat{a})$ and $q=(1/\sqrt{2})(\hat{a}^{\dagger}+\hat{a})$. 

We now consider two different cases: when $\alpha=0$ and when $\alpha\neq0$. 

\textbullet\, If $\alpha=0$, Eq. \eqref{eq:clasicoRM} corresponds to the usual RM. Furthermore, if $\lambda>\lambda_{c}$, in the superradiant phase, the critical points $(p^{*},q^{*})\in\mathbb{R}^{2}$, satisfying $\eval{\nabla H_{\alpha=0}}_{(p^{*},q^{*})}=0$, are \cite{Puebla2016} 

\begin{equation}\label{eq:criticalpoints}
(p^{*},q^{*})\in\left\{(0,0),\,\,\,\frac{1}{2\lambda\omega}\left(0,\sqrt{8\lambda^{4}-\frac{\omega^{2}\omega_{0}^{2}}{2}}\right),\,\,\,\frac{1}{2\lambda\omega}\left(0,-\sqrt{8\lambda^{4}-\frac{\omega^{2}\omega_{0}^{2}}{2}}\right)\right\}.
\end{equation}
Note that since $H_{0}(p,q)=H_{0}(p,-q)$, the second and third critical points correspond to the exact same energy, the ground-state energy, while the point $(0,0)$ corresponds to the ESQPT critical energy, $H_{0}(0,0)=-1$. 
If $\lambda<\lambda_{c}$, the only critical point is $(0,0)$, corresponding to the ground-state in the normal phase.

In Fig. 2(a) of the main text we show seven contour lines, Eq. \eqref{eq:clasicoRM}, for the RM with $\lambda=3\lambda_c$, in the superradiant phase. As in the main text, we set $\omega=1$ and $\omega_{0}=300$. We first note that Eq. \eqref{eq:clasicoRM} indeed gives rise to two equivalent, symmetrical minima, corresponding to a degenerate ground-state.  We observe that below the critical energy of the ESQPT, $\epsilon_{c}=-1$, these contour lines are split into two disjoint parts: one in which $q<0$, and another one in which $q>0$. By contrast, above the critical energy there is just one contour line covering both positive and negative values of $q$. This means that any initial condition with $E<E_c$ gives rise to a trajectory $\mathbf{x}(t)=(p[t],q[t])$ in which the sign of the coordinate $q(t)$ remains always positive or negative, depending on its initial value, $q(0)$. Contrarily, in any trajectory with $E>E_c$ the sign of $q(t)$ changes as a function of time, i.e., it does not remain constant. Thus, we observe that the dynamical function $f(\mathbf{x}[t])$ that allows to distinguish excited-state phases is particularly simple in this case: $f(t)=q(t)$. Then, $\textrm{sign}\,q(t)$ is indeed a classical constant of motion in the region with $E<E_{c}$, and otherwise it is not. Taking into account that the bosonic quadrature is $q=(\hat{a}^{\dagger}+\hat{a})/\sqrt{2}$, one arrives at 
\begin{equation}\label{eq:constanterabialpha0}\hatmath{C}=\textrm{sign}\,(\hat{a}^{\dagger}+\hat{a}),\end{equation} 
which is a particular case of the expression given in the main text. 

\textbullet \, If $\alpha\neq0$, the previous constant energy curves are increasingly distorted as $|\alpha|$ grows. Importantly, as soon as $\alpha\neq0$, Eq. \eqref{eq:clasicoRM} ceases to have two equivalent minima; instead, the ground-state loses its previous degeneracy and there appears a second local energy minimum, which is asymmetrical with respect to the global energy minimum (the ground-state). Unfortunately, it appears that obtaining a general analytical expression of the critical points when $\alpha\neq0$ may not be a straightforward matter. 

In Fig. 2(d) of the main text we show the contour plots of such \textit{perturbed} energies with $\alpha=1/\sqrt{3}$. For the chosen value of the model parameters, the critical points and the corresponding reduced energies are $\epsilon_{\textrm{GS}}[p=0,q=q^{*}(\alpha=1/\sqrt{3},\lambda=\frac{3}{2}\sqrt{\omega\omega_{0}})=-46.6022]=-7.3265$, $\epsilon_{c2}[0,q=q^{*}(\alpha=1/\sqrt{3},\lambda=\frac{3}{2}\sqrt{\omega\omega_{0}})=26.2078]=-2.46034$, and $\epsilon_{c}[0,q=q^{*}(\alpha=1/\sqrt{3},\lambda=\frac{3}{2}\sqrt{\omega\omega_{0}})=1.3224]=-0.9572$. The first one, $\epsilon_{\textrm{GS}}$, is the ground-state energy; the second, $\epsilon_{c_{2}}$, is another local minimum, distinct from the first, which gives rise to an ESQPT characterized by an abrupt jump in the density of states, as it is shown in Fig. 2(e) of the main text; and the third, $\epsilon_{c}$, is the critical energy of the same ESQPT that appears in the case with $\alpha=0$, characterized by a logarithmic singularity in the density of states, as it is shown in Figs. 2(b) and 2(e) of the main text. It is noteworthy that even though the curves are severely distorted with respect to those of Fig. 2(a), a completely analogous argument can be made for the separation of the phase space and the definition of the dynamical function $f(\mathbf{x}[t])$. The difference is that when $\alpha\neq0$, the ESQPT curves do not intersect at $q_{c}(\alpha=0,\lambda>3\lambda_{c})=0$, but at a certain value $q_{c}(\alpha,\lambda)>0$ ($\alpha>0$) or $q_{c}(\alpha,\lambda)<0$ ($\alpha<0$) instead; here, this happens at $q_{c}(\alpha=1/\sqrt{3},\lambda=\frac{3}{2}\sqrt{\omega\omega_{0}})=1.3324$. Therefore, the proposed constant of motion also holds if one suitably rewrites it: ${f}(t)={q}(t)-{q}_{c}(\alpha,\lambda)$, and thus for this deformed, asymmetrical variant of the RM, the quantum version of the constant reads

\begin{equation}\hat{\mathcal{C}}=\textrm{sign}\,[(\hat{a}^{\dagger}+\hat{a})/\sqrt{2}-1.3224],\end{equation}
which is a particular case of Eq. (2) in the main text when $q_{c}(\alpha=1/\sqrt{3},\lambda=3\lambda_{c})=1.3224$. 

\subsection{Dicke model}

The semiclassical phase space of the DM is more involved because it has two effective degrees of freedom ($\nu=2$), but it shows the same behavior regarding the bosonic coordinate $q$ \cite{Bastarrachea2014}. Indeed, using the Glauber-Bloch coherent states the semiclassical constant energy surfaces can be shown to equal (see, e.g., \cite{Pilatowsky2021})

\begin{equation}
\label{eq:clasicoDM}
H_{\alpha}(p,q;P,Q)\equiv \frac{\widetilde{H}_{\alpha}(p,q;P,Q)}{\omega_{0}j}=\frac{\omega}{2\omega_{0}}(p^{2}+q^{2})+\frac{1}{2}(P^{2}+Q^{2})+\frac{2\lambda q Q}{\omega_{0}}\sqrt{1-\frac{P^{2}+Q^{2}}{4}}-1+\sqrt{\frac{N}{\omega_{0}j}}\alpha q,
\end{equation}
where the parameters have the same physical meaning as in the RM. Here, the phase space coordinates $(p,q)$ have been rescaled as $p=(i/\sqrt{2j})(\hat{a}^{\dagger}-\hat{a})$ and $q=(1/\sqrt{2j})(\hat{a}^{\dagger}+\hat{a})$, and the creation and annihilation operators as a function of $(p,q)$ are $\hat{a}=\sqrt{j/2}(q+ip)$ and $\hat{a}^{\dagger}=\sqrt{j/2}(q-ip)$.  Therefore it is clear that in this case the semiclassical phase space is four-dimensional, $\mathcal{M}\subseteq \mathbb{R}^{4}$.By appropriately projecting these energy surfaces, an equivalent argument to the case of the RM can be made. We briefly comment on the critical points obtained from this Hamiltonian. 

\textbullet\, If $\alpha=0$, Eq. \eqref{eq:clasicoDM} is the standard DM. If $\lambda>\lambda_{c}$, its extrema are known analytically. They are of the form $(p^{*},q^{*};P^{*},Q^{*})=(0,q^{*};0,Q^{*})$ where 
\begin{equation}\label{eq:critDM}
(q^{*},Q^{*})\in \left\{(0,0),\left(-\sqrt{\frac{4\lambda^{2}}{\omega^{2}}-\frac{\omega_{0}^{2}}{4\lambda^{2}}},\sqrt{2-\frac{\omega\omega_{0}}{2\lambda^{2}}}\right),\left(\sqrt{\frac{4\lambda^{2}}{\omega^{2}}-\frac{\omega_{0}^{2}}{4\lambda^{2}}},-\sqrt{2-\frac{\omega\omega_{0}}{2\lambda^{2}}}\right)\right\}.
\end{equation}
Since $H_{0}(p,q;P,Q)=H_{0}(p,-q;P,-Q)$, the second and third extrema correspond to the same energy, the ground state energy, while the first extrema corresponds to the ESQPT critical point, with energy $H_{0}(0,0,0,0)=-1$. If $\lambda<\lambda_{c}$, the only extremum of $H_{0}$ is again $H_{0}(0,0,0,0)=-1$, which is now the ground-state. The constant of motion in this case is the same as in Eq. \eqref{eq:constanterabialpha0}. 

\textbullet\, If $\alpha\neq0$, the situation becomes more complicated, as in the RM above. As a case study, we focus on $\alpha=1/2$, as in the main text. For this perturbation strength, and setting $\omega=\omega_{0}=1$, we obtain the following extrema and associated energies. In all cases, $p=P=0$. The ground-state energy is $\epsilon_{\textrm{GS}}[q=q^{*}(\alpha=1/2,\lambda=\frac{3}{2}\sqrt{\omega\omega_{0}})=-3.69497,Q=1.34919]=-6.91626$. The second local minimum occurs at $\epsilon_{c_{2}}[q=q^{*}(\alpha=1/2,\lambda=\frac{3}{2}\sqrt{\omega\omega_{0}})=2.26081,Q=-1.30701]=-2.70149$, and corresponds to an ESQPT characterized by an abrupt jump in the derivative of the density of states. Finally, the critical energy of the same ESQPT that appears in the case with $\alpha=0$, characterized by a logarithmic singularity in the derivative of the density of states, has been shifted to $\epsilon_{c}[q=q_{c}(\alpha=1/2,\lambda=\frac{3}{2}\sqrt{\omega\omega_{0}})=0.0921344,Q=-0.268854]=-0.968103.$ Therefore in this case, the constant of motion is written

\begin{equation}
    \hat{\mathcal{C}}=\textrm{sign}[(\hat{a}^{\dagger}+\hat{a})/\sqrt{2j}-0.0921344],
\end{equation}
which is again a particular case of Eq. (2) in the main text. 

\section{S2. Comparison of the variance of $\hat{\mathcal{C}}$ with a smaller size}

In the main text we show that $\sigma_{\mathcal{C}}^{2}$ displays a jump at the ESQPT critical energy where $\hat{\mathcal{C}}$ ceases to be a constant of motion [for additional information about this quantity, see. Sec. S5]. Below this energy, $\sigma_{\mathcal{C}}^{2}$ is zero for both models; above this energy, it is non-zero. We mentioned that this jump is not so abrupt in the DM as in the RM because the system-size parameter, $\mathcal{N}$, is smaller in the DM than it is in the RM. The reason for this is that diagonalizing the DM is much more computationally expensive than the RM, and we cannot approach the thermodynamic limit further than we have. This jump should only be sharp strictly in the thermodynamic limit. 

\begin{center}
\begin{figure}[h!]
\hspace*{-0cm}\includegraphics[width=0.6\textwidth]{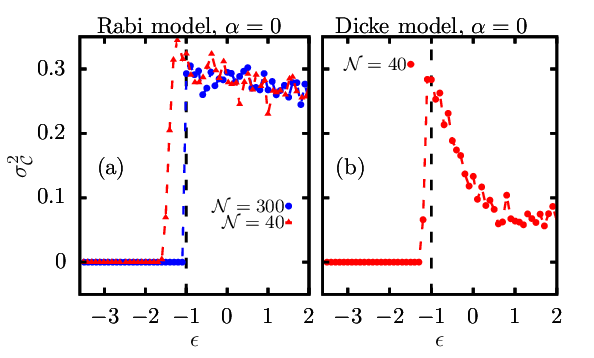} 
\caption{Variance $\sigma_{\mathcal{C}}^{2}$ as a function of the reduced energy $\epsilon$, for (a) the Rabi model, and two different values of the system-size parameter, $\mathcal{N}=40,300$, and for (b) the Dicke model, for $\mathcal{N}=40$, as in the main text ($\lambda=3\lambda_{c}$). }
\label{fig:constantereferee}
\end{figure}
\end{center}

In Fig. \ref{fig:constantereferee} we show, for the case $\alpha=0$, the same curves as in Fig. 1 of the main text for the RM and DM (similar results can be obtained for $\alpha\neq0$). Additionally, we plot the same quantity for $\mathcal{N}=40$ in the RM. We can see in red the curves for the same system-size $\mathcal{N}=40$, while the case $\mathcal{N}=300$ of the RM, used in the main text, is in blue here. We observe that if $\mathcal{N}=40$, the jump in $\sigma_{\mathcal{C}}^{2}$ is also less abrupt in the RM, too. Thus, this is compatible with the less abrupt jump observed in the DM in the main text: the reason is that the parameter leading to the thermodynamic limit, $\mathcal{N}$, is smaller in the DM than it is in the RM. This result is fully compatible with the finite-size scaling shown in Fig. 3 of the main text, which is the same for both the RM and the DM.

We also mentioned in the main text that there are some differences in the behavior and actual value of $\sigma_{\mathcal{C}}^{2}$ in both models once the ESQPT is crossed. It seems plausible that this is a consequence of the difference in degrees of freedom of the RM ($\nu=1$) and the DM ($\nu=2$). We also suspect that quantum chaos may play a role here, as it develops in the DM at high energies; however, there is no quantum chaos in the RM because it has a single degree of freedom (i.e., it does not allow for spectral fluctuations). 

\section{S3. Some features of the operator $\hat{\mathcal{C}}$}
Here we comment on two aspects of $\hat{\mathcal{C}}$ in the RM and DM with $\alpha=0$, whose details we left unattended in the main text. 

\textbullet\, \textbf{$\hat{\mathcal{C}}$ changes parity of any Fock state.} Let us study how $\hat{\mathcal{C}}$ acts on the eigenstates of $\hat{\Pi}$. To do so, it is useful to make use of the integral representation \cite{fn1,Roberts1980}
\begin{equation}\label{eq:integral}\hat{\mathcal{C}}=\frac{2}{\pi}\int_{0}^{\infty}\textrm{d}x\,\hat{f}(x^{2}\mathbb{I}+\hat{f}^{2})^{-1}, \end{equation} 
where $\hat{f}$ is the quantum version of the dynamical function, $f$, allowing to separate the phase space trajectories, as indicated in the main text. For example, in the unperturbed RM and DM ($\alpha=0)$, $\hat{f}=\hat{a}^{\dagger} + \hat{a}$. First, it is clear that $\hat{f}$ changes the parity of Fock states, as it either increases or decreases the number of photons by one, keeping constant the value of $j_{z}$.  This implies that $\hat{f}^{2}$ commutes with parity, $[\hat{f}^{2},\hat{\Pi}]=0$. In general, one has that $[\hat{f}^{2n},\hat{\Pi}]=0$ for all $n\in\mathbb{N}$, while $\hat{f}^{2n-1}$ changes the parity. We now have that since $x^{2}\mathbb{I}+\hat{f}^{2}$ commutes with parity, for all $x$, then $(x^{2}\mathbb{I}+\hat{f}^{2})^{-1}$ also commutes with parity. Therefore, $\hat{f}(x^{2}\mathbb{I}+\hat{f}^{2})^{-1}$ must change the parity of any Fock state, for all $x$, and using the above integral representation, Eq. \eqref{eq:integral}, it follows that $\hatmath{C}$ also changes the parity of any state, which is what we wished to prove. 

This actually predicts the appearance of energy doublets (degeneracies) in the spectrum below the ESQPT critical energy, $E_{n}<E_{c}$. Interpreted as subspace applications, we have that  $\hat{\Pi}:\mathcal{E}_{n}\to\mathcal{E}_{n}$, where $\mathcal{E}_{n}$ is the subspace with energy $E_{n}$, and also  $\hat{\mathcal{C}}:\mathcal{E}_{n}\to\mathcal{E}_{n}$ whenever $E_{n}<E_{c}$, in the TL, as suggested in the finite-size scaling presented in Fig. 3 of the main text. Since $\hat{\mathcal{C}}$ changes parity as showed above, there are two distinct eigenbases diagonalizing the Hamiltonian, $\{\ket{E_{n,\pm}}\}$ and $\{\ket{\varepsilon_{n,\pm}}\}$, that is,

\begin{equation}\label{bases}
\textrm{(i)}\,\,\,\hat{\mathcal{H}}_{0}\ket{E_{n,\pm}}=E_{n}\ket{E_{n,\pm}},\,\,\,\hat{\Pi}\ket{E_{n,\pm}}=\pm\ket{E_{n,\pm}},\,\,\,\textrm{and}\,\,\,\,\,\,\textrm{(ii)}\,\,\,\, \hat{\mathcal{H}}_{0}\ket{\varepsilon_{n,\pm}}=E_{n}\ket{\varepsilon_{n,\pm}},\,\,\,\,\,\,\hat{\mathcal{C}}\ket{\varepsilon_{n,\pm}}=\pm\ket{\varepsilon_{n,\pm}}.
\end{equation}
This can only happen if eigenlevels are doubly degenerate below the ESQPT critical energy, if $E<E_{c}$.

\textbullet\, \textbf{Eigenstates and eigenvalues of $\hat{\mathcal{C}}$.} Let us calculate the eigenstates of $\hat{\mathcal{C}}$ in the low-energy phase, $E<E_{c}$, where it is a constant of motion. To do so, we consider a single eigenspace, $\mathcal{E}_{n}$, and we work in the parity eigenbasis, $\hat{\Pi}\ket{E_{n,\pm}}=\pm\ket{E_{n,\pm}}$. As $\hat{\mathcal{C}}$ is a constant of motion, both of its eigenstates must be linear combinations of $\ket{E_{n,+}}$ and $\ket{E_{n,-}}$ in this $n$th eigenspace; in other words, we must have $\hat{\mathcal{C}}(\gamma\ket{E_{n,+}}+\delta\ket{E_{n,-}})=(\gamma\ket{E_{n,+}}+\delta\ket{E_{n,-}})$ and also $\hat{\mathcal{C}}(\delta\ket{E_{n,+}}-\gamma\ket{E_{n,-}})=-(\delta\ket{E_{n,+}}-\gamma\ket{E_{n,-}}$, where $|\gamma|^{2}+|\delta|^{2}=1$. Furthermore, as $\hat{\mathcal{C}}$ changes parity, $\hat{\mathcal{C}}\ket{E_{n,+}}=\alpha\ket{E_{n,-}}$ and $\hat{\mathcal{C}}\ket{E_{n,-}}=\beta\ket{E_{n,+}}$ for some $\alpha,\beta\in\mathbb{C}$. Because $\hat{\mathcal{C}}$ is a real operator, the only solutions to all of these conditions are (i) $(\alpha=\beta=1,\gamma=\delta=1/\sqrt{2})$ and (ii) $(\alpha=\beta=-1,\gamma=\delta=1/\sqrt{2})$. Both give rise to the same pair of eigenvectors of $\hat{\mathcal{C}}$, namely $\ket{\varepsilon_{n,\pm}}\equiv (\ket{E_{n,+}}\pm \ket{E_{n,-}})/\sqrt{2}$. The eigenvalues of $\hat{\mathcal{C}}$ are therefore only $+1$ or $-1$, and as we see their precise sign ($+$ or $-$) is only defined up to the relative phase between the Hamiltonian eigenstates. 

\section{S4. Finite-size scaling: additional figures}
\begin{center}
\begin{figure}[h!]
\hspace*{-1cm}\includegraphics[width=0.7\textwidth]{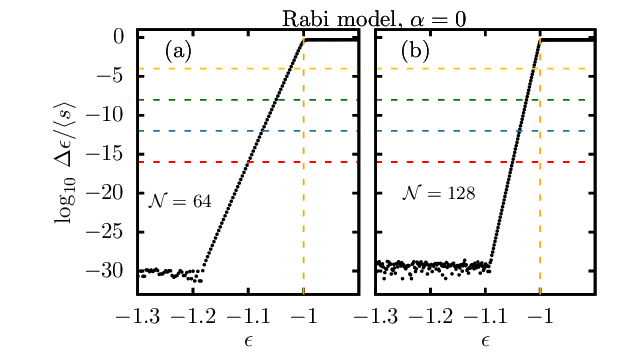} 
\caption{Normalized energy gap of eigenstates of the RM ($\alpha=0$) with opposite parity, $\Delta \epsilon_{n}=| \epsilon_{n,+}-\epsilon_{n,-}|/\langle s\rangle$ (black points), where $\langle s\rangle$ is the mean level spacing calculated within a window of 10 eigenlevels around the target energy. Panel (a): $\mathcal{N}=64$; panel (b): $\mathcal{N}=128$. The horizontal dashed lines represent the bounds $\gamma=10^{-16},10^{-12},10^{-8},10^{-4}$ (from bottom to top) used in the main text. The orange vertical dashed line represents the ESQPT critical energy, $\epsilon_{c}=-1$. Results are given in quadruple precision (128-bit arithmetic). }
\label{fig:scalingreferee}
\end{figure}
\end{center}

For simplicity, in this section we focus on the RM (the DM produces equivalent results). In Fig. \ref{fig:scalingreferee} we show how to obtain the finite-size precursor of the ESQPT critical energy, denoted $\varepsilon(\mathcal{N},\gamma)$ in the main text. We plot the gap between energies associated to eigenstates of opposite parity as a function of the reduced energy $\epsilon$ for two different system sizes: $\mathcal{N}=64$ in (a), and $\mathcal{N}=128$ in (b). The horizontal dashed lines mark the various arbitrary bounds $\gamma$ of the main text (same color code as in Fig. 3 of the main text), and the vertical dashed line marks the ESQPT critical energy in the thermodynamic limit as obtained from the semiclassical analogue, $\epsilon_{c}=-1$. Thus, all (normalized) energy gaps below a given $\gamma$ are smaller than $\gamma$, $\Delta\epsilon/\langle s\rangle<\gamma$. For a given $\mathcal{N}$ and $\gamma$, the ESQPT energy precursor of finite systems, $\varepsilon (\mathcal{N},\gamma)$, is calculated as the energy, $\epsilon$, at which $\Delta \epsilon/\langle s\rangle$ crosses the horizontal line of that particular $\gamma$. We observe that as we increase the thermodynamic limit parameter, $\mathcal{N}$, if we fix a given bound $\gamma$, the precursor $\varepsilon(\mathcal{N},\gamma)$ gets closer to the thermodynamic limit ESQPT energy, $\epsilon_{c}=-1$. In other words, the degeneracy is preserved up to larger energies as the system size is increased. We also observe that if we fix $\mathcal{N}$ and vary the bound $\gamma$, the energy precursor gets closer to $\epsilon_{c}=-1$ as $\gamma$ is increased from $10^{-16}$ to $10^{-4}$; this is simply because the arbitrary bound is less strict if $\gamma$ is larger. It is interesting to note that in the transition region (the `ramp' in $\Delta\epsilon/\langle s\rangle$) occupies a smaller width in energy and is sharper as $\mathcal{N}$ increases. Fig. 3 of the main text provides numerical support to the conjecture that this transition becomes sharp (the `ramp' gets vertical) strictly in the TL, $\mathcal{N}\to\infty$.

\begin{center}
\begin{figure}[h!]
\hspace*{-1cm}\includegraphics[width=0.7\textwidth]{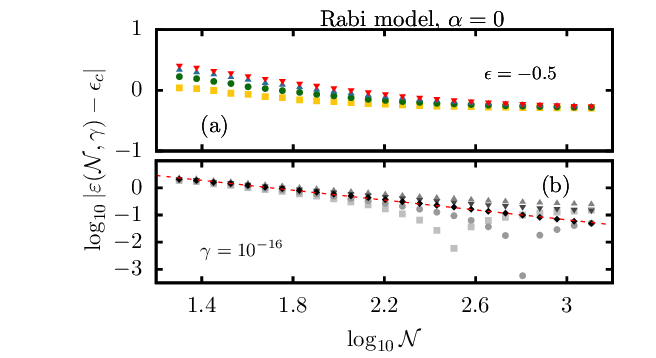} 
\caption{Finite-size scaling of the ESQPT precursor $\varepsilon(\mathcal{N},\gamma)$ as a function of $\mathcal{N}$. In (a) we fix $\epsilon=-0.5$ and we vary $\gamma=10^{-16},10^{-12},10^{-8},10^{-4}$ (from top to bottom). In (b) $\gamma=10^{-16}$ is fixed and the energies $\epsilon=-1.2,-1.1,-0.8,-0.9,-1$ are considered (in this order, from gray to dark black).  }
\label{fig:scalingsreferee3}
\end{figure}
\end{center}

Additionally, in Fig. \ref{fig:scalingsreferee3} we present two modifications of the finite-size scaling of Fig. 3 of the main text. In panel (a) we show the scaling for a fixed energy, $\epsilon$, which is different than the actual ESQPT critical energy obtained from the semiclassical analysis, $\epsilon_{c}$, and we let the bounds $\gamma$ vary, exactly as in the main text. In particular, we choose $\epsilon_{c}\to\epsilon=-0.5$. We clearly observe that the scaling behavior $|\varepsilon(\mathcal{N},\gamma)-\epsilon_{c}|\propto\mathcal{N}^{-\beta}$ is obviously no longer fulfilled: beyond a given $\mathcal{N}$, all curves show a plateau. In panel (b) we fix $\gamma=10^{-16}$ instead, and we change the value of $\epsilon$ to which we compare the ESQPT energy precursor $\varepsilon(\mathcal{N},\gamma)$. Besides $\epsilon_{c}=-1$, we also choose two other energies below and above it ($\epsilon=-1.2,-1.1$ and $\epsilon=-0.9,-0.8$). The red dashed line represents the best linear fit (in log-log scale) to the points for $\epsilon=\epsilon_{c}=-1$. We observe that $\epsilon=-1$ is the only case for which the scaling behavior $|\varepsilon(\mathcal{N},\gamma)-\epsilon_{c}|\propto\mathcal{N}^{-\beta}$ is fulfilled. The behavior of the curves with $\epsilon>\epsilon_{c}$ differ from those with $\epsilon<\epsilon_{c}$. When $\epsilon>\epsilon_{c}$, we get smooth plateaus in the curves, which bend beyond a given $\mathcal{N}$ and thus do not follow the finite-size scaling proposed as $\mathcal{N}\to\infty$. When $\epsilon<\epsilon_{c}$, we find quite sharp dips at a given $\mathcal{N}$ for each $\epsilon$; this is because at those points the precursor $\varepsilon(\mathcal{N},\gamma)$ almost matches that value of $\epsilon$, and thus the logarithm of its difference naturally tends to $-\infty$. 

These results, together with those in Fig. 3 of the main text, strongly suggest that in the thermodynamic limit, $\mathcal{N}\to\infty$, the eigenlevel degeneracy should break down exactly at the ESQPT critical energy, $\epsilon_{c}=-1$. This phenomenon is accompanied by the fact that $\hat{\mathcal{C}}$ ceases to be a constant of motion at that exact energy too, in the thermodynamic limit. As $\mathcal{N}$ is increased, the indicators presented get closer to this asymptotic behavior, $\mathcal{N}\to\infty$. 

\section*{S5. Time evolution of the operator $\hat{\mathcal{C}}$}\label{secs5}
\begin{center}
\begin{figure}[h!]
\hspace*{-0cm}\includegraphics[width=0.45\textwidth]{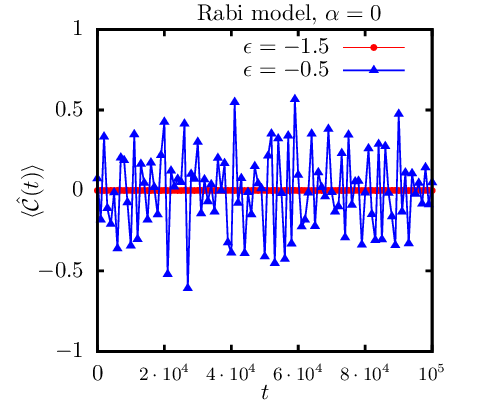}
\caption{Time evolution $\langle\hat{\mathcal{C}}(t)\rangle$ in the RM ($\alpha=0$). Initial states are chosen with energy below the ESQPT ($\epsilon=-1.5$) and above the ESQPT ($\epsilon=-0.5$). Below the ESQPT, $\langle\hat{\mathcal{C}}(t)\rangle$ is a constant, while above the ESQPT it fluctuates around some equilibrium value.  }
\label{fig:timeevo}
\end{figure}
\end{center}

One of the conclusions of the manuscript is that below the ESQPT critical energy, the operator $\hat{\mathcal{C}}$ is a constant, meaning that its expectation values, $\langle\hat{\mathcal{C}}(t)\rangle\equiv \bra{\Psi(t)}\hat{\mathcal{C}}\ket{\Psi(t)}$, in eigenstates with energy $\epsilon<\epsilon_{c}=-1$, do not change over time. By contrast, this is not the case above the ESQPT, $\epsilon>\epsilon_{c}$, where $\hat{\mathcal{C}}$ is no longer constant, so its expectation values show fluctuations around its long-time equilibrium value, $\overline{\langle\hatmath{C}\rangle}$, that depends on the initial condition.  In Fig. \ref{fig:timeevo} we illustrate these facts with two simple cases, considering the RM with $\alpha=0$ for definiteness. We have selected the initial state $\ket{\Psi(t=0)}$ such that $\langle\hatmath{C}\rangle=0$ below the critical energy. This is accomplished by picking an homogoneous superposition of 6 energy eigenspaces -- i.e., the probability of obtaining any one of them in an energy measurement is exactly $1/6$. Three terms have been chosen of the form $(1/\sqrt{2})[\ket{E_{n,+}}+\ket{E_{n,-}}]$, and the remaining three terms of the form $(1/\sqrt{2})[\ket{E_{n,+}}-\ket{E_{n,-}}]$, i.e., the relative phase is the opposite. The initial state $\ket{\Psi(t=0)}$ is obtained by simply adding together all these terms and normalizing the resulting state so that $\bra{\Psi(0)}\ket{\Psi(0)}=1$. Above the ESQPT, $\ket{\Psi(t=0)}$ is composed by 12 consecutive eigenstates with the same coefficients as in the former case.

The quantity that we calculate in Fig. 1 of the main text is the variance of the expectation value of $\hatmath{C}$ in a given state $\ket{\Psi(t)}$ evolving in time. Formally, 
\begin{equation}\label{sigma}
    \sigma_{\mathcal{C}}^{2}\equiv \lim_{\tau\to\infty}\frac{1}{\tau}\int_{0}^{\tau}\textrm{d}t\,\left[\langle \hatmath{C}(t)\rangle-\overline{\langle\hatmath{C}\rangle}\right]^{2},
\end{equation}
where $\overline{\langle\hatmath{C}\rangle}=\lim_{\tau\to\infty}(1/\tau)\int_{0}^{\tau}\textrm{d}t\,\bra{\Psi(t)}\hatmath{C}\ket{\Psi(t)}$ is the usual long-time expectation value of $\hatmath{C}$ in the vector state $\ket{\Psi(t)}$ and $\langle\hatmath{C}(t)\rangle=\bra{\Psi(t)}\hatmath{C}\ket{\Psi(t)}$ is the expectation value of $\hatmath{C}$ in the state $\ket{\Psi(t)}$ at time $t$. As we mention in the main text, the content of our conjecture is that below the ESQPT $\hatmath{C}$ is a constant of motion. Then its expectation value does not change over time, that is, $\sigma_{\mathcal{C}}^{2}=0$ for all states $\ket{\Psi(t)}$. 

We remark that Eq. \eqref{sigma} is in general different from the variance (dispersion) in the measurement of $\hatmath{C}$ in a given, fixed time, $t$, \begin{equation}\label{sigmar}
\Delta_{\mathcal{C}}^{2}(t)\equiv \bra{\Psi(t)} (\hatmath{C}-\langle\hatmath{C}(t)\rangle)^{2}\ket{\Psi(t)}=\bra{\Psi(t)}\hatmath{C}^{2}\ket{\Psi(t)}-(\bra{\Psi(t)}\hatmath{C}\ket{\Psi(t)})^{2}=\langle\hatmath{C}^{2}(t)\rangle-\langle\hatmath{C}(t)\rangle^{2}.
\end{equation}
Eqs. \eqref{sigma} and \eqref{sigmar} are not equal; importantly, $\sigma_{\mathcal{C}}^{2}$ does not depend on time but $\Delta_{\mathcal{C}}^{2}(t)$ does. Still, there are some implications between them which hold true. If $\Delta_{\mathcal{C}}^{2}(t)=0$ for all $t$, then $\sigma_{\mathcal{C}}^{2}=0$ also; it follows that if $\sigma_{\mathcal{C}}^{2}\neq0$, then $\Delta_{\mathcal{C}}^{2}(t)\neq0$ for all $t$ as well. However, if $\Delta_{\mathcal{C}}^{2}(t)\neq0$ for a given $t$, it does not follow that $\sigma_{\mathcal{C}}^{2}\neq0$. This is illustrated in Fig. \ref{fig:timeevo} by the curve for $\epsilon=-1.5<\epsilon_{c}$, which has $\Delta_{\mathcal{C}}^{2}(t)=1$ for all $t$ ($\langle\hatmath{C}^{2}(t)\rangle=1$ by definition, and $\langle\hatmath{C}(t)\rangle$ is always zero in this case), but $\sigma_{\mathcal{C}}^{2}=0$ as the curve clearly shows (there are no fluctuations in time, $\langle\hatmath{C}(t)\rangle=\overline{\langle\hatmath{C}\rangle}$ for all $t$). This means that the time-evolved state, $\ket{\Psi(t)}$, remains a balanced superposition of the two eigenstates of $\hatmath{C}$, corresponding to its eigenvalues $+1$ and $-1$, for all times. In other words, the time evolution conserves the cat-like structure --- a balanced superposition of the two classical wells --- of the initial state. The curve for $\epsilon=-0.5>\epsilon_{c}$ corresponds to $\Delta_{\mathcal{C}}^{2}(t)\neq 1$ in general (as the figure shows, $\langle\hatmath{C}(t)\rangle$ is in general not zero and can deviate quite significantly), and $\sigma_{\mathcal{C}}^{2}\neq0$ (there are fluctuations in time around the long-time average $\overline{\langle\hatmath{C}\rangle}$).

\section*{S6. Initial states used and quench} 
For Fig. 1 of the main text we mention that we use as initial states 10 consecutive eigenstates of the corresponding Hamiltonian centered at a given energy and with the same population. This means that we fix the desired target energy, $\epsilon$, and look for the eigenstate of the RM or DM closest in energy to $\epsilon$, $\bra{\epsilon}\hatmath{H}\ket{\epsilon}=\epsilon$. Then we pick a total of 10 eigenstates above and below $\epsilon$ and the initial state that we use at time $t=0$ is a superposition of these eigenstates with equal population, $\ket{\Psi(t=0)}=\sum_{i}c_{i}\ket{E_{i}}$ with $c_{i}=1/\sqrt{10}$. This protocol is repeated changing the target energy $\epsilon$ to probe different parts of the spectrum.

\begin{center}
\begin{figure}[h!]
\hspace*{-0cm}\includegraphics[width=0.55\textwidth]{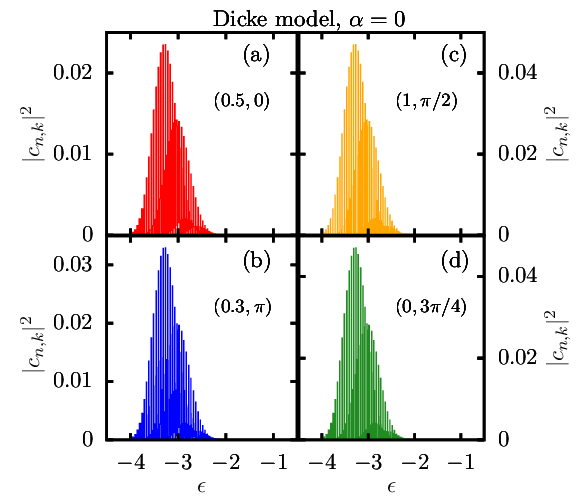}
\caption{Coeficients $|c_{n,k}|^{2}$ as a function of the energy $\epsilon$ after a quench $\lambda_{\textrm{ini}}=\frac{3}{2}\lambda_{c}\to\lambda_{\textrm{fin}}=3\lambda_{c}$ for different values of the pair $(p,\phi)$. Results are for the Dicke model ($\alpha=0$).}
\label{fig:coefs}
\end{figure}
\end{center}

For Fig. 4 in the main text, we perform a quench in the control parameter of the Hamiltonian, $\lambda_{\textrm{ini}}=3/2\lambda_{c}\to\lambda_{\textrm{fin}}=3\lambda_{c}$. In this case the initial state is a general superposition of the (degenerate) ground-state of the Dicke model at $\lambda_{\textrm{ini}}$, $\ket{E_{\textrm{GS},+}}$ and $\ket{E_{\textrm{GS},-}}$: $\ket{\Psi(t=0)}=\sqrt{p}\ket{E_{\textrm{GS},+}}+e^{i\phi}\sqrt{1-p}\ket{E_{\textrm{GS},-}}$. This superposition depends on an amplitude, $p$, and an angle that fixes the overall phase, $\phi$. 
Therefore, before the quench the eigenstates $\ket{E_{\textrm{GS},\pm}}$, of the same energy, $E_{\textrm{GS}}$, and different parity, $\pm$, may or may not be equally populated depending on $p$. In any case, only those two eigenstates are populated. After the quench, the initial state evolves in the quenched Hamiltonian and this population changes: a certain range of eigenstates are populated, not just the two initial states. This population, $|c_{n,k}|^{2}$, is shown for different values of $(p,\phi)$ in Fig. \ref{fig:coefs} as a function of the energy. The average energy is approximately $\epsilon_{\textrm{fin}}\approx -3.15<\epsilon_{c}=-1$ in all cases, as indicated in the main text. We observe that the quench does not populate all eigenstates of the final Hamiltonian between the minimum and maximum energy for which $|c_{n,k}|^{2}$ is significantly different from zero. This is because at low energies there exists an approximate integral of motion in the Dicke Hamiltonian that gives rise to independent energy bands. Thus, depending on the initially populated eigenstates some bands are not accessible after the quench, and the eigenstantes corresponding to those bands are not populated. This property was described in Refs. \cite{Bastarrachea2017,Relano2016}.

